
\input tex_inputs:maclyon
\magnification\magstep1
\baselineskip = 0.5 true cm

  \def\sa{\vskip 0.30 true cm}
  \def\sb{\vskip 0.60 true cm}
  \def\sc{\vskip 0.15 true cm}
  \def\sd{\vskip 0.50 true cm}
  \def\demi{ { {\lower 3 pt\hbox{$\scriptstyle 1$}} \over
               {\raise 3 pt\hbox{$\scriptstyle 2$}} } }

  \pageno = 1
  \vsize = 25.9   true cm
  \hsize = 15.65 true cm
  \voffset = -1 true cm

\font\msim=msym10
\def\gr{\hbox{\msim R}}

\def\grn{\hbox{\msim N}}

\rightline{LYCEN 9242}
\rightline{October 1992}

\sc
\sb
\sb

\centerline {\bf SOME ASPECTS OF $q$-BOSON
CALCULUS\footnote* {\sevenrm Contribution to the Symposium
``Symmetries in Science VI: From the Rotation Group to Quantum
Algebras'' held at the Cloister Mehrerau (Bregenz, Austria, 2-7
August 1992) in honor of Professor L.C.~Biedenharn on the
occasion of his 70th birthday.
Published in the
Proceedings of
the Symposium,
ed.~B.~Gruber
(Plenum Press, New York),
pp.~691-704.}}

\vskip 0.4 true cm

\sa
\sb

\vskip 0.5 true cm

\centerline {Yu.F.~Smirnov\footnote* {\sevenrm
Permanent address~: Institute of Nuclear Physics,
Moscow State University, 119899 Moscow, Russia.}
and M.R.~Kibler}

\sa

\centerline {Institut de Physique Nucl\'eaire de Lyon}
\centerline {IN2P3-CNRS et Universit\'e Claude Bernard}
\centerline {43 Boulevard du 11 Novembre 1918}
\centerline {F-69622 Villeurbanne Cedex}
\centerline {France}

\sa

\sa
\sa
\sb
\sb

\baselineskip = 0.7 true cm

  \sa

\vfill\eject

\baselineskip = 0.5 true cm

\vglue 5 true cm

\noindent {\bf SOME ASPECTS OF $q$-BOSON CALCULUS}

\vskip 0.4 true cm

\sa
\sb

\vskip 0.5 true cm

\leftskip = 1.6 true cm

{Yu.F.~Smirnov\footnote* {\sevenrm
Permanent address~: Institute of Nuclear Physics,
Moscow State University, 119899 Moscow, Russia.}
and M.R.~Kibler}

\sa

{Institut de Physique Nucl\'eaire de Lyon}

{IN2P3-CNRS et Universit\'e Claude Bernard}

{43 Boulevard du 11 Novembre 1918}

{F-69622 Villeurbanne Cedex}

{France}

\leftskip = 0 true cm

\sa
\sb
\sa
\sa
\sb
\sb

\baselineskip = 0.5 true cm

\noindent {\bf 1. PRELIMINARIES}

\sb

The content of this work is concerned with Jordan-Schwinger
calculus, using deformed bosons, that has been so largely
investigated by Biedenharn. It is thus a real pleasure to
dedicate this work to Larry Biedenharn on the occasion of his
70th birthday. The works by Biedenharn and by Biedenharn and
his collaborators have been a source of inspiration for both
authors of the present paper, especially in connection with (i) the
Wigner-Racah algebra of a chain of compact groups (involving finite
groups) in view of its applications to nuclear, molecular and
condensed matter physics and, more recently, with (ii) quantum
algebras.

This work constitutes a first step towards a complete study of
the $su_q(2)$ unit tensor. The components $t[q : k \rho \Delta]$
of such a tensor operator are defined by$^{1,2}$
$$
< j' m' \vert t[q : k \rho \Delta] \vert j m > =
\delta(j', j + \Delta)
\delta(m', m + \rho  ) (-1)^{2k} [2j'+1]^{-{1\over2}}
 (j k m \rho \vert j' m')_q
\eqno (1)
$$
where
$(j k m \rho \vert j' m')_q$ is a Clebsch-Gordan coefficient
(CGc) for $su_q(2)$
as defined, for example, in ref.~3.
(In eq.~(1) and in the following, we employ the usual notation
$[a] = (q^a - q^{-a}) / (q - q^{-1})$ with $q \in \gr$.)
The operator $t[q : k \rho \Delta]$
constitutes a $q$-deformation of the operator
$t_{k q \alpha} \equiv t[1 : k q \alpha]$ worked out in ref.~2.
Our program
is to find a realization of $t[q : k \rho \Delta]$ in terms of
$q$-bosons. We address here the first part of this program
by defining in section 2 the ($q$-deformed) Schwinger algebra
relative to $su_q(2)$ and by giving in section 3 an algorithm for
producing recurrent relations (RR's) between CGc's for $su_q(2)$.

\sb
\sd

\noindent {\bf 2. THE $q$-DEFORMED SCHWINGER ALGEBRA}

\sb

It is well known that the quantum algebra $su_q(2)$, which is
characterized (in the Kulish-Reshetikhin-Drinfeld-Jimbo
realization) by
$$
[J_3, J_-] \; = \; - \; J_- \qquad
[J_3, J_+] \; = \; + \; J_+ \qquad
[J_+, J_-] \; = \; [2J_3]
\eqno (2)
$$
(plus other axioms relative to its Hopf algebraic
structure), may be realized by means of two commuting sets of
$q$-bosons. In fact, the generators $J_-$, $J_3$ and $J_+$ of
$su_q(2)$ can be written as
$$
J_- \; = \; a^+_- a_+ \qquad
J_3 \; = \; {1 \over 2} \left( N_1 - N_2 \right) \qquad
J_+ \; = \; a^+_+ a_-
\eqno (3)
$$
in terms of the $q$-deformed annihilation ($a_+$ and $a_-$), creation
($a_+^+ \equiv (a_+)^{\dagger}$ and
 $a_-^+ \equiv (a_-)^{\dagger}$) and number
($N_1 = (N_1)^{\dagger}$ and
 $N_2 = (N_2)^{\dagger}$)
operators. These operators satisfy$^{4,5}$
$$
\eqalign{
 [a_+, a_-]     \; = \;
 [a^+_+, a^+_-] \; = \;
 [a_+, a^+_-]   \;&= \;
 [a^+_+, a_-]   \; = \; 0          \cr
 [N_1,a^+_+] =  a^+_+ \quad
 [N_1,a  _+] = -a  _+ \quad
 [N_2,a^+_-]&=  a^+_- \quad
 [N_2,a  _-] = -a  _-              \cr
 a^+_+ a_+ \; = \; [N_1]     \quad
 a_+ a^+_+ \; = \; [N_1 + 1] \quad
 a^+_- a_- \; =&\; [N_2]     \quad
 a_- a^+_- \; = \; [N_2 + 1]       \cr
}
\eqno (4)
$$
In a (two-particle) Fock space ${\cal F}_1 \otimes {\cal F}_2$,
with the basis vectors
$$
\vert n_1 n_2> = { 1 \over \sqrt{[n_1]![n_2]!} } \>
(a^+_+)^{n_1}
(a^+_-)^{n_2} \> \vert 00>
\eqno (5)
$$
(where $[n]!$ stands for the $q$-deformed factorial),
the $a$'s and $N$'s act in the following way
$$\eqalign{
  a_+   \; |n_1n_2> \; = \; &{\sqrt {[n_1]}}\; |n_1 - 1, n_2>\cr
  a_+^+ \; |n_1n_2> \; = \; &{\sqrt {[n_1 + 1]}}\; |n_1 + 1, n_2>\cr
  a_-   \; |n_1n_2> \; = \; &{\sqrt {[n_2]}}\; |n_1, n_2 - 1>\cr
  a_-^+ \; |n_1n_2> \; = \; &{\sqrt {[n_2 + 1]}}\; |n_1, n_2 + 1>\cr
      N_i |n_1n_2 > \; = \; &n_i |n_1n_2> \qquad (i=1,2)
  }
\eqno (6)
$$
from which it is clear that
$a_+^+ = (a_+)^{\dagger}$ and
$a_-^+ = (a_-)^{\dagger}$ when $q \in \gr$ (as
supposed in this paper)   or   $q \in S^1$.

If the $su(2)$ notations are introduced for the basis vectors (5), namely,
$$
     |jm> \, \equiv \, |n_1n_2>                    \qquad
             j \, = \, {1 \over 2} (n_1 + n_2)     \qquad
             m \, = \, {1 \over 2} (n_1 - n_2)
\eqno (7)
$$
then, the operators $J_-$, $J_3$ and $J_+$ (to be considered,
in physical applications, as $q$-analogues of spherical angular
momentum operators) act on ${\cal F}_1 \otimes {\cal F}_2$
through
$$
\eqalign{
  J_- \; |jm > \; = \; &{\sqrt {[j + m] \; [j - m + 1]}} \; |j, m-1 >\cr
  J_3 \; |jm > \; = \; &m \; |jm >\cr
  J_+ \; |jm > \; = \; &{\sqrt {[j - m] \; [j + m + 1]}} \; |j, m+1 >\cr
}
\eqno (8)
$$
showing that the vectors (5) with a fixed value $2j$ $(2j \in \grn)$ of
$n_1+n_2$ span the irrep ($j$) of the quantum algebra $su_q(2)$.

Like for $su_q(2)$, the quantum algebra
         $su_q(1,1) \simeq sp_q(2,\gr)$ can be
realized in terms of the two commuting sets
$\left\{ a_+, a_+^+ \right\}$ and
$\left\{ a_-, a_-^+ \right\}$. This algebra is generated by the
operators
$$
K_- \; = \; a_+a_- \qquad
K_3 \; = \; {1 \over 2} \; (N_1 + N_2 + 1) \qquad
K_+ \; = \; a^+_+ a^+_-
\eqno (9)
$$
which satisfy the commutation relations
$$
[K_3, K_-] \; = \; - \; K_- \qquad
[K_3, K_+] \; = \; + \; K_+ \qquad
[K_+, K_-] \; = \; - \; [2K_3]
\eqno (10)
$$
that are typical of $su_q(1,1)$. The generators $K_-$, $K_3$
and $K_+$ act on ${\cal F}_1 \otimes {\cal F}_2$ according to
$$
\eqalign{
  K_- \; |jm > \; = \; &{\sqrt {[j-m] \, [j+m]}}     \; |j-1, m >\cr
  K_3 \; |jm > \; = \; &(j+{1 \over 2}) \; |jm >\cr
  K_+ \; |jm > \; = \; &{\sqrt {[j-m+1] \, [j+m+1]}} \; |j+1, m >\cr
}
\eqno (11)
$$
and may thus be considered as $q$-analogues of
hyperbolical angular momentum operators.

The algebras $su_q(2)$ and $su_q(1,1)$ do not commute. The four
nonvanishing commutators of the $J$'s and $K$'s may serve to
define new bilinear forms of the $a$'s. Indeed, by introducing
the operators
$$
k^+_+ \; = \; - \; a^+_+ a^+_+ \qquad
k^+_- \; =      \; a^+_- a^+_- \qquad
k^-_- \; = \; - \; a_+ a_+     \qquad
k^-_+ \; =      \; a_- a_-
\eqno (12)
$$
we can put the nonvanishing commutators of type $[J,K]$ in the
form
$$\matrix{
  [J_+, K_+] & =
 &k^+_+ ( [ K_3 - J_3 - {1\over 2} ] &-&
          [ K_3 - J_3 + {1\over 2} ] )
 \cr
  [J_+, K_-] & =
 &k_+^- ( [ K_3 + J_3 - {1\over 2} ] &-&
          [ K_3 + J_3 + {1\over 2} ] )
 \cr
  [J_-, K_+] & =
 &k^+_- ( [ K_3 + J_3 + {1\over 2} ] &-&
          [ K_3 + J_3 - {1\over 2} ] )
 \cr
  [J_-, K_-] & =
 &k^-_- ( [ K_3 - J_3 + {1\over 2} ] &-&
          [ K_3 - J_3 - {1\over 2} ] )
 \cr
}
\eqno (13)
$$
which go to $-k^+_+$, $-k^-_+$, $+k^+_-$, $+k^-_-$, respectively,
in the limiting case $q=1$. The $k$'s are step operators in
the space ${\cal F}_1 \otimes {\cal F}_2$ since we have
$$
\eqalign{
  k^+_+ \; |jm > \; = \; &- {\sqrt {[j+m+1] \, [j+m+2]}} \; |j+1, m+1 >\cr
  k^+_- \; |jm > \; = \; &+ {\sqrt {[j-m+1] \, [j-m+2]}} \; |j+1, m-1 >\cr
  k^-_- \; |jm > \; = \; &- {\sqrt {[j+m-1] \, [j+m]}}   \; |j-1, m-1 >\cr
  k^-_+ \; |jm > \; = \; &+ {\sqrt {[j-m-1] \, [j-m]}}   \; |j-1, m+1 >\cr
}
\eqno (14)
$$
It is clear that the operators $\left\{ J,K,k \right\}$ close under
commutation in the limiting case $q=1$. In this case, they
generates the 10-dimensional noncompact Lie algebra
$sp(4, \gr) \simeq so(3,2)$. In the case $q \ne 1$,
the operators $\left\{ J,K,k \right\}$ span the quantized
universal enveloping algebra $U_q(so(3,2)) \equiv so_q(3,2)
                                           \simeq sp_q(4,\gr)$
that we shall refer to as the $q$-deformed Schwinger algebra. (This
terminology follows from the angular momentum context of ref.~2.
In another context, $so_q(3,2)$ may be called a $q$-deformed de
Sitter algebra.) The nonvanishing commutators for $so_q(3,2)$, besides
(2), (10) and (13), are given in ref.~6 and reproduced in the appendix.
Note that a realization of the algebra $so_q(3,2)$ in the
Bargmann-Fock space may be found by making the replacements
$$
\eqalign{
a^+_+     &\mapsto z_1                          \qquad
a  _+      \mapsto D_{z_1}                        \qquad
N_1        \mapsto z_1 { \partial \over {\partial z_1} } \cr
a^+_-     &\mapsto z_2                          \qquad
a  _-      \mapsto D_{z_2}                        \qquad
N_2        \mapsto z_2 { \partial \over {\partial z_2} } \cr
}
\eqno (15)
$$
where the finite difference operator $D_{x}$ defined via
$$
D_{x} f(x) = { {f(qx) - f(q^{-1}x)} \over {(q-q^{-1})x} }
\eqno (16)
$$
is the Jackson derivative.

At this point, we should discuss the Hopf algebraic structure
of $su_q(2)$, $su_q(1,1)$ and $so_q(3,2)$. This is well
documented for $su_q(2)$ and $su_q(1,1)$
(see, for instance, refs.~4 and 7).
For the algebra $so_q(3,2)$, a coproduct, a
counit and an antipode should be defined in order to endow
this algebra with an Hopf algebraic structure. Indeed, the
generators of the quantized de Sitter algebra $so_q(3,2)$
have been given, in a Cartan-Weyl basis, and the Hopf
algebraic structure of $so_{q}(3,2)$ has been discussed
explicitly in ref.~8.
The passage formulas, which shall be reported elsewhere,
between the generators of $so_q(3,2)$ in a
Cartan-Weyl basis and the operators $J$'s, $K$'s and $k$'s
allows us to consider the $q$-deformed Schwinger algebra as an
Hopf algebra.

The noncommutativity of $su_q(2)$ and $su_q(1,1)$ seems to
prevent the irrep's of both algebras to be fixed
simultaneously. However, the second order invariant operators
$$
C_2(su_q(2))   = J_- J_+ + [J_3] [J_3 + 1]
\eqno (17)
$$
and
$$
C_2(su_q(1,1)) =-K_+ K_- + [K_3] [K_3 - 1]
\eqno (18)
$$
can be diagonalized simultaneously.
Indeed, eqs.~(17) and (18) can be expressed as
$$
C_2(su_q(2))    = [K_3]^2 - [{1 \over 2}]^2
\equiv [C_2(u_q(1))]^2    - [{1 \over 2}]^2
\eqno (19)
$$
and
$$
C_2(su_q(1,1))  = [J_3]^2 - [{1 \over 2}]^2
\equiv [C_2(so_q(2))]^2   - [{1 \over 2}]^2
\eqno (20)
$$
(with evident definitions of $u_q(1)$ and $so_q(2)$), respectively.
Equations (19) and (20) are valid as far as matrix elements,
on the Fock space ${\cal F}_1 \otimes {\cal F}_2$, are
concerned. The invariants of $su_q(2)$ and $u_q(1)$, on one
hand, and of $su_q(1,1)$ and $so_q(2)$, on the other hand,
are thus connected. Furthermore, $J_3$
commutes with the three generators of $su_q(1,1)$~:
$$
[J_3, K_{\nu}] = 0 \qquad \nu = {-},3,{+}
\eqno (21)
$$
while $K_3$ commutes with the three generators of $su_q(2)$~:
$$
[K_3, J_{\nu}] = 0 \qquad \nu = {-},3,{+}
\eqno (22)
$$
According to the definition given in ref.~9, from eqs.~(19)-(22)
we deduce that the algebra $su_q(1,1)$, generated by the
set $\left \{ K_-, K_3, K_+ \right \}$, and the algebra
$so_q(2)$, generated by the operator $J_3$, are complementary
in the frame of some definite representation of the host
algebra $sp_q(4,\gr)$.
Similarly,
the algebra $su_q(2)$, generated by the
set $\left \{ J_-, J_3, J_+ \right \}$, and the algebra
$u_q(1)$, generated by the operator $K_3$, are complementary
within some representation of $sp_q(4,\gr)$.
Indeed, two chains are relevant here,
viz.,
$$
\eqalign{
sp_q(4,\gr) \simeq so_q(3,2) \supset & sp_q(2,\gr) \simeq su_q^k(1,1)
                             \supset u_q^{\kappa}(1) \cr
sp_q(4,\gr) \simeq so_q(3,2) \supset & su_q^j(2)
                             \supset so_q^{m}(2)     \cr
}
\eqno (23)
$$
for which we have two pairs of complementary algebras~:
$(su_q^k(1,1), so_q^{m}(2))$ and $(su_q^j(2), u_q^{\kappa}(1))$.
The symbols $j$, $m$, $k$ and $\kappa$ labelling the irrep's of
the algebras $su_q(2)$, $so_q(2)$, $su_q(1,1)$ and $u_q(1)$,
respectively, are put as superscripts. We conclude that the
vector
$$
\vert n_1 n_2 > \equiv \vert j m > \equiv \vert k \kappa >
\eqno (24)
$$
can be considered simultaneously as (i) a basis vector for the
irrep $(j)$ of $su_q(2)$, in an $su_q(2) \supset so_q(2)$ basis,
with spherical angular momentum
$$
j = {1 \over 2} (n_1 + n_2)
\eqno (25)
$$
and 3-axis projection (eigenvalue of $J_3$)
$$
m = {1 \over 2} (n_1 - n_2)
\eqno (26)
$$
and as (ii) a basis vector for the
irrep $(k{+})$, belonging to the positive discrete series of
$su_q(1,1)$, in an $su_q(1,1) \supset u_q(1)$ basis,
with hyperbolical angular momentum
$$
k = {1 \over 2} (n_1 - n_2 -1) = m - {1 \over 2}
\eqno (27)
$$
and 3-axis projection (eigenvalue of $K_3$)
$$
\kappa  = {1 \over 2} (n_1 + n_2 + 1) = j + {1 \over 2}
\eqno (28)
$$
Note that, in the $su_q(1,1)$ notations, eq.~(11) can be
rewritten in the useful form
$$
\eqalign{
  K_- \; |k \kappa > \; = \; &{\sqrt {[\kappa + k] \, [\kappa - k -1]}} \;
|k , \kappa - 1 >\cr
  K_3 \; |k \kappa > \; = \; & \kappa \; |k \kappa >\cr
  K_+ \; |k \kappa > \; = \; &{\sqrt {[\kappa - k] \, [\kappa + k +1]}} \;
|k , \kappa + 1 >\cr
}
\eqno (29)
$$
The vectors $\vert k \kappa >$ and $\vert j m >$ are
eigenvectors of the invariant operators $C_2(su_q(1,1))$ and
$C_2(su_q(2))$ with the eigenvalues $[k][k+1]$ and
$[j][j+1]$, respectively. For the host algebra
$sp_q(4,\gr) \simeq so_q(3,2)$, only two infinite-dimensional
irrep's, namely, $[\dot 0]$ and $[\dot 1]$, belonging to the
positive discrete series, are realized in the Fock space
${\cal F}_1 \otimes {\cal F}_2$ (relative to a two-dimensional
$q$-oscillator). The basis
vectors (5) with $n_1 + n_2$ even belong to
$[\dot 0]$ and the ones with $n_1 + n_2$ odd to $[\dot 1]$.

To close this section, let us briefly discuss the irreducible
tensor character of the $q$-bosons $a_+$, $a_+^+$, $a_-$ and
$a_-^+$, a problem also addressed by Biedenharn and
Tarlini,$^{10}$ Nomura$^{11,12}$ and Quesne.$^{13}$
First, we note that eq.~(6) yields
$$
\eqalign{
a_+ \; |jm> \; = \; & {\sqrt{[j + m]}}\; |j - {1\over 2}, m - {1\over 2}>\cr
a_+^+ \; |jm> \; = \; & {\sqrt{[j + m+1]}}\; |j + {1\over 2}, m + {1\over
2}>\cr
a_- \; |jm> \; = \; & {\sqrt{[j - m]}}\; |j - {1\over 2}, m + {1\over 2}>\cr
a_-^+ \; |jm> \; = \; & {\sqrt{[j - m + 1]}}\; |j + {1\over 2}, m - {1\over 2}>
\cr
}
\eqno (30)
$$
in the $su_q(2)$ notations. Equation (30) shows that the sets
$\left \{ a_+^+, a_-^+ \right\}$ and
$\left \{ a_-  , a_+   \right\}$
are connected to the $su_q(2)$ irreducible tensorial sets
$$
\left \{ t[q:\demi,\rho, \demi]:\rho=\pm \demi \right\}
\quad {\rm and} \quad
\left \{ t[q:\demi,\rho,-\demi]:\rho=\pm \demi \right\},
$$
respectively. Both sets transform as the
unitary irrep $({1 \over 2})$ of $su_q(2)$.
Second, in the $su_q(1,1)$ notations, we have
$$
\eqalign{
a_+ \; |k \kappa>
\; = \; & {\sqrt{[\kappa + k]}}\;   |k - {1\over 2}, \kappa - {1\over 2}>\cr
a_+^+ \; |k \kappa>
\; = \; & {\sqrt{[\kappa + k+1]}}\; |k + {1\over 2}, \kappa + {1\over 2}>\cr
a_- \; |k \kappa>
\; = \; & {\sqrt{[\kappa - k-1]}}\; |k + {1\over 2}, \kappa - {1\over 2}>\cr
a_-^+ \; |k \kappa>
\; = \; & {\sqrt{[\kappa - k]}}\;   |k - {1\over 2}, \kappa + {1\over 2}>
\cr
}
\eqno (31)
$$
which indicate that the sets
$\left \{ a_+^+, a_-   \right\}$ and
$\left \{ a_-^+, a_+   \right\}$
are connected to the $su_q(1,1)$ irreducible tensorial sets
$$
\left\{ \tau[q:\demi,\rho, \demi]:\rho=\pm \demi\right\}
\quad {\rm and} \quad
\left\{ \tau[q:\demi,\rho,-\demi]:\rho=\pm \demi\right\},
$$
respectively. (The operator $\tau[q : k \rho \Delta]$ for $su_q(1,1)$
parallels      the operator $t   [q : k \rho \Delta]$ for $su_q(2)$.)
The latter sets have well defined transformation properties
with respect to the nonunitary irrep
$({1 \over 2})$ of $su_q(1,1)$.
As a conclusion, the four $q$-boson operators $a_+$, $a_+^+$, $a_-$ and $a_-^+$
can be united in a double unit tensor operator
$w(q)^{{1 \over 2}{1 \over 2}}$, with
components $w(q)^{{1 \over 2}{1 \over 2}}_{{\rho}{\sigma}}$ where
$\rho  =\pm{1 \over2}$ labels its components with respect to $su_q(2)$ and
$\sigma=\pm{1 \over2}$ does the same         with respect to $su_q(1,1)$.
By applying the Wigner-Eckart theorem for both algebras, we have
$$
\eqalign{
< j' m'      \vert w(q)^{{1 \over 2}{1 \over 2}}_{{\rho}{\sigma}}
\vert j m>      & = a(j) \> (j \demi m      \rho   \vert j' m'     )_q
\cr
< k' \kappa' \vert w(q)^{{1 \over 2}{1 \over 2}}_{{\rho}{\sigma}}
\vert k \kappa> & = b(k) \> (k \demi \kappa \sigma \vert k' \kappa')_q
}
\eqno (32)
$$
Equation (32) leads to some relations, to be discussed
elsewhere, connecting $su_q(2)$ CGc's of type
$(j {1 \over 2} m      \rho   \vert j' m'     )_q$
and $su_q(1,1)$ CGc's of type
$(k {1 \over 2} \kappa \sigma \vert k' \kappa')_q$. The extension of
these relations through the use of double tensors of higher rank would be
interesting.

\sb
\sd

\noindent {\bf 3. TOWARDS RECURRENT RELATIONS FOR $su_q(2)$}

\sb

\noindent {\bf 3.1. The Case $q=1$}

\sb

The main ingredient of the method developed in
ref.~2 is the use of four commuting pairs
$\left\{ a_+, a_+^+ \right\}$,
$\left\{ a_-, a_-^+ \right\}$,
$\left\{ b_+, b_+^+ \right\}$ and
$\left\{ b_-, b_-^+ \right\}$ of ordinary bosons. The $a$'s and
$b$'s serve for constructing two copies of $su^j(2)$ (say,
$su^{j_1}(2)$ and $su^{j_2}(2)$), respectively. (The Lie algebra
$su^j(2)$ is defined by (2) with $q=1$.) In other words, the
 two-dimensional harmonic oscillator (with Fock space
${\cal F}_1 \otimes {\cal F}_2$ or
${\cal F}_3 \otimes {\cal F}_4$) of section 2 is replaced by a
four-dimensional harmonic oscillator
(with Fock space
${\cal F} = {\cal F}_1 \otimes {\cal F}_2 \otimes
            {\cal F}_3 \otimes {\cal F}_4$).
Therefore, eq.~(5) with $q=1$ yields
$$
\vert j_1 m_1> = { 1 \over \sqrt{(j_1+m_1)!(j_1-m_1)!} } \>
(a^+_+)^{j_1+m_1}
(a^+_-)^{j_1-m_1} \> \vert 00>
\eqno (33)
$$
and
$$
\vert j_2 m_2> = { 1 \over \sqrt{(j_2+m_2)!(j_2-m_2)!} } \>
(b^+_+)^{j_2+m_2}
(b^+_-)^{j_2-m_2} \> \vert 00>
\eqno (34)
$$
 in $su^{j_1}(2)$ (with Fock space ${\cal F}_1 \otimes {\cal F}_2)$
and $su^{j_2}(2)$ (with Fock space ${\cal F}_3 \otimes {\cal F}_4)$
notations, respectively.

Three commuting Lie algebras, denoted here as $su^{\cal J}(2)$,
$su^{\Lambda}(2)$ and $su^{\cal K}(1,1)$, come to play an important
role in ref.~2. The generators of $su^{\cal J}(2)$ are
$$
\eqalign{
{\cal J}_3 & = {1 \over 2} (N_1 + N_2 - N_3 - N_4) \cr
{\cal J}_+ & = a^+_+ b_+  +  a^+_- b_-             \cr
{\cal J}_- & = b^+_+ a_+  +  b^+_- a_-             \cr
}
\eqno (35)
$$
where $N_1 = a^+_+ a_+$,
      $N_2 = a^+_- a_-$,
      $N_3 = b^+_+ b_+$ and
      $N_4 = b^+_- b_-$. The algebra $su^{\Lambda}(2)$
corresponds to the sum of the spherical angular momenta
associated to $su^{j_1}(2)$ and $su^{j_2}(2)$ in the sense
that $su^{\Lambda}(2)$ is generated by
$$
\eqalign{
{\Lambda}_3 & = {1 \over 2} (N_1 - N_2 + N_3 - N_4) \equiv (J_1)_3 + (J_2)_3\cr
{\Lambda}_+ & = a^+_+ a_-  +  b^+_+ b_-             \equiv (J_1)_+ + (J_2)_+\cr
{\Lambda}_- & = a^+_- a_+  +  b^+_- b_+             \equiv (J_1)_- + (J_2)_-\cr
}
\eqno (36)
$$
Finally, the algebra $su^{\cal K}(1,1)$ is spanned by the
operators
$$
\eqalign{
{\cal K}_3 & = {1 \over 2} (N_1 + N_2 + N_3 + N_4) + 1 \cr
{\cal K}_+ & = a^+_+ b^+_-   -   a^+_- b^+_+           \cr
{\cal K}_- & = a_+   b_-     -   a_-   b_+             \cr
}
\eqno (37)
$$
It can be easily verified that the three algebras
$su^{\cal J}(2)$, $su^{\Lambda}(2)$ and $su^{\cal K}(1,1)$
commute. Another important point arises from the fact
that the eigenvalues of the Casimir operators
$C_2(su^{\cal J}  (2))$,
$C_2(su^{\Lambda} (2))$ and
$C_2(su^{\cal K}(1,1))$ are all equal, say to $j(j+1)$~;
therefore, the irrep's of the three algebras may be labelled by
a common (quantum) number $j$. (This number refers to a
spherical, or compact,
angular momentum for $su^{\cal J}(2)$ and $su^{\Lambda}(2)$
and to an hyperbolical, or noncompact,
angular momentum for $su^{\cal K}(1,1)$.)
The commutativity of the algebras
$su^{\cal J}(2)$, $su^{\Lambda}(2)$ and $su^{\cal K}(1,1)$
together with the coincidence of the spectra of their Casimir
operators constitute an evidence for complementarity relations, in the
sense of Moshinsky and Quesne,$^9$ between
these algebras.

In order to better understand these complementarity relations,
let us consider the more general case of $mn$ pairs of boson
operators corresponding to an $mn$-dimensional harmonic
oscillator. Two chains of Lie groups (or Lie algebras) may be
exhibited in this case~:
$$
\eqalign{
Sp(2mn,\gr) \supset & Sp(2m,\gr) \supset  U(m) \supset \cdots
\cr
Sp(2mn,\gr) \supset & U(n)       \supset SO(n) \supset \cdots
\cr
}
\eqno (38)
$$
for which there are two pairs of complementary groups, namely,
$(Sp(2m,\gr),SO(n))$ and $(U(m),U(n))$. The invariant operators of
the groups in a given pair are connected in a simple manner. We
deal here with a four-dimensional oscillator and
the situation $mn=4$ presents some specificities. In this
situation, the chains (38) may be specialized as
$$
\eqalign{
Sp(8,\gr) \supset & Sp(2,\gr) \simeq SU^{\cal K}(1,1) \supset U^{\kappa}(1)
\cr
Sp(8,\gr) \supset & U(4)      \supset SO(4) \simeq
SU^{\cal J}    (2) \otimes SU^{\Lambda}    (2) \supset
SO^{M_{\cal J}}(2) \otimes SO^{M_{\Lambda}}(2)
\cr
}
\eqno (39)
$$
for which the two relevant pairs of complementary groups are
$(SU^{\cal K}(1,1) , SO(4))$ and
$(U            (1) , U (4))$.
The breaking of $SO(4)$ into
$SU^{\cal J}(2) \otimes SU^{\Lambda}(2)$
leads indeed to the lucky situation for which we have three complementary
groups~: $SU^{\cal J}(2)$, $SU^{\Lambda}(2)$ and
$SU^{\cal K}(1,1)$. The latter three groups correspond
to the three complementary Lie
algebras defined by eqs.~(35)-(37).

To go further within the just decribed complementarity
relations, some precisions are in order. We know that only the
symmetric irrep's $< n >$ of $u(4)$ can be realized
in the Fock space ${\cal F}$ of a four-dimensional oscillator
($< n >$ denotes the Young diagram associated to the
total number $n=n_1 + n_2 + n_3 + n_4$ of quanta).
Furthermore, only the irrep's $(\omega,0)$, of class I in the
terminology of Vilenkin, of the algebra $so(4)$ are realized in
the space ${\cal F}$. Let us write the two second-order
invariants of $so(4)$ in the form
$$
C_2(so(4)) = 2 (J^2 + {\Lambda}^2) \qquad
C_2(so(4))'= 2 (J^2 - {\Lambda}^2)
\eqno (40)
$$
where $J^2 = C_2(su^{\cal J }(2))$ and
$\Lambda^2 = C_2(su^{\Lambda}(2))$. The two commuting algebras
$su^{\cal J }(2)$ and $su^{\Lambda}(2)$ are complementary in
the framework of the irrep's $(\omega,0)$ of class I of the
algebra $so(4)$. The Casimir operators $J^2$ and $\Lambda^2$
have the same eigenvalues $j(j+1)$. To go
from the $so(4)$ to the
$su^{\cal J}(2) \oplus su^{\Lambda}(2)$ notations, we have to
use (cf.~ref.~14)
$$
{\cal J} = {\Lambda} \equiv j = {1 \over 2} \omega
\eqno (41)
$$
so that the eigenvalues of $C_2(so(4))$ and $C_2(so(4))'$ are
$\omega(\omega + 2)$ and $0$, respectively. The complementarity
of $so(4)$ and $sp(2,\gr) \simeq su^{\cal K}(1,1)$ manifests
itself by the commutativity of $su^{\cal K}(1,1)$ with both
$su^{\cal J}(2)$ and $su^{\Lambda}(2)$ and by the fact that the
Casimir operator $C_2(su^{\cal K}(1,1))$ has the eigenvalues
$j(j+1)$.

As a result, we can construct basis vectors
$\vert n \omega M_{\cal J} M_{\Lambda}>$ associated to the chain
$u(4) \supset so(4) \simeq su^{\cal J}(2) \oplus su^{\Lambda}(2)
      \supset so^{M_{\cal J }}(2) \oplus
              so^{M_{\Lambda}}(2)$
and having well defined transformation properties with respect
to the three algebras
$su^{\cal J}(2)$, $su^{\Lambda}(2)$ and $su^{\cal K}(1,1)$. It
is appropriate to introduce the notation
$$
\vert n \omega M_{\cal J} M_{\Lambda}> \equiv \vert j : \mu m \kappa>
\eqno (42)
$$
where $\mu$, $m$ and $\kappa$ are the ``projections'' of the
``angular momentum'' $j = {1 \over 2} \omega$ onto the
``directions'' $so^{M_{\cal J }}(2)$,
               $so^{M_{\Lambda}}(2)$ and
$u^{\kappa}(1)$, respectively. More precisely, we shall have
$$
\eqalign{
\mu   & = M_{\cal J}  = j_1 - j_2  = {1 \over 2} (n_1 + n_2 - n_3 - n_4)  \cr
  m   & = M_{\Lambda} = m_1 + m_2  = {1 \over 2} (n_1 - n_2 + n_3 - n_4)  \cr
\kappa&=                j_1 + j_2+1= {1 \over 2} (n_1 + n_2 + n_3 + n_4)+1\cr
}
\eqno (43)
$$
as eigenvalues of the operators ${\cal J}_3$, ${\Lambda}_3$ and
${\cal K}_3$.

The vector (42) is obtained from a linear combination of the
vectors $\vert n_1 n_2 n_3 n_4 >$. In turn, the
vector  $\vert n_1 n_2 n_3 n_4 >$ in the Fock space
${\cal F}$ can be identified to $\vert j_1 m_1> \otimes \>
                                 \vert j_2 m_2>$
when considered as basis vector for the tensor product
$(j_1) \otimes
 (j_2)$, in $su^{\cal J}(2)$, of the irrep's
$(j_1)$ and
$(j_2)$ with
$$
2j_1 = n_1 + n_2 \quad 2m_1 = n_1 - n_2 \quad
2j_2 = n_3 + n_4 \quad 2m_2 = n_3 - n_4 \quad
\eqno (44)
$$
In a similar way, $\vert n_1 n_2 n_3 n_4>$ can be
simultaneously considered as a basis vector
$\vert \lambda_1 \mu_1> \otimes \>
 \vert \lambda_2 \mu_2>$
for the tensor product
$(\lambda_1) \otimes
 (\lambda_2)$, in $su^{\Lambda}(2)$, of the irrep's
$(\lambda_1)$ and
$(\lambda_2)$ with
$$
2\lambda_1 = n_1 + n_3 \quad 2\mu_1 = n_1 - n_3 \quad
2\lambda_2 = n_2 + n_4 \quad 2\mu_2 = n_2 - n_4 \quad
\eqno (45)
$$
Finally, the vector $\vert n_1 n_2 n_3 n_4>$ can be
also considered as a basis vector
$\vert k_1 \kappa_1> \otimes \>
 \vert k_2 \kappa_2>$
for the tensor product
$(k_1+) \otimes
 (k_2+)$, in $su^{\cal K}(1,1)$, of the irrep's
$(k_1+)$ and
$(k_2+)$ with
$$
2k_1 = n_1 - n_2 - 1 \quad 2\kappa_1 = n_1 + n_2 + 1 \quad
2k_2 = n_3 - n_4 - 1 \quad 2\kappa_2 = n_3 + n_4 + 1 \quad
\eqno (46)
$$
The coupled basis vector
$$
\vert j_1 j_2 jm > = \sum_{m_1m_2} (j_1 j_2 m_1 m_2 \vert jm) \;
\vert j_1 m_1 > \otimes \> \vert j_2 m_2>
\eqno (47)
$$
for the decomposition of $(j_1) \otimes (j_2)$ into
$\bigoplus_{\vert j_1 - j_2 \vert}^{j_1 + j_2} (j)$
should coincide with the vector
$ \vert n \omega M_{\cal J} M_{\Lambda} > $ because both of
them are labelled unambiguously by four numbers
$$
\eqalign{
          n & =    2 (j_1 + j_2)=              n_1 + n_2 + n_3 + n_4
\quad \omega = 2j
\cr
M_{\cal J } & = \mu = j_1 - j_2 = {1 \over 2} (n_1 + n_2 - n_3 - n_4)
\cr
M_{\Lambda} & =   m = m_1 + m_2 = {1 \over 2} (n_1 - n_2 + n_3 - n_4)
\cr
}
\eqno (48)
$$
Therefore, eq.~(42) can be reinterpreted as
$$
\vert j_1 j_2 j m > \equiv
\vert j : \mu m \kappa>
\eqno (49)
$$
with the projections
$\mu = j_1 - j_2$, $m$ and $\kappa = j_1 + j_2 + 1$. Similarly,
the coupled basis vector $ \vert \lambda_1 \lambda_2 j \mu> $
for the
decomposition of $(\lambda_1) \otimes (\lambda_2)$ into
$\bigoplus_{\vert \lambda_1 - \lambda_2 \vert}^{\lambda_1 + \lambda_2} (j)$
can be taken as
$$
\vert \lambda_1 \lambda_2 j \mu > \equiv
\vert j : \mu m \kappa>
\eqno (50)
$$
with the projections
$\mu$, $m=\lambda_1 - \lambda_2$ and $\kappa = \lambda_1 + \lambda_2 + 1$.

The passage for the vectors $\vert j : \mu m \kappa>$
from the $su^{\cal J }(2)$ notation (49) to the
         $su^{\Lambda}(2)$ notation (50) makes it possible to generate
several families of useful relations between the
CGc's for the group $SU(2)$. As a matter of fact,
the inner scalar product
$< n_1 n_2 n_3 n_4 \vert j : \mu m \kappa >$ may be
tackled in two ways~:
$$
 < n_1 n_2 n_3 n_4 \vert j : \mu m \kappa > =
(      j_1       j_2   m_1   m_2 \vert j   m) \ \, {\rm or} \ \,
(\lambda_1 \lambda_2 \mu_1 \mu_2 \vert j \mu)
\eqno (51)
$$
which provide the key of an algorithm
for generating various relations between
$SU(2)$ CGc's.

As a first example, by taking eqs.~(44) and (45) into account, the
identity (51) between
$(      j_1       j_2   m_1   m_2 \vert j   m)$ and
$(\lambda_1 \lambda_2 \mu_1 \mu_2 \vert j \mu)$ yields
$$
\eqalign{
 (j_1 j_2 m_1 m_2 \vert j m) = & \cr
       (&{{j_1 + m_1 + j_2 + m_2} \over 2} \>
         {{j_1 - m_1 + j_2 - m_2} \over 2} \cr
        &{{j_1 + m_1 - j_2 - m_2} \over 2} \>
         {{j_1 - m_1 - j_2 + m_2} \over 2} \vert j j_1 - j_2
       ) \cr
}
\eqno (52)
$$
which is nothing but a Regge symmetry property.

A second family of relations concerns the derivation of three-
and four-term RR's. The possibility to replace
the scalar product
$< n_1 n_2 n_3 n_4 \vert j : \mu m \kappa >$ by one
of the two CGc's of (51) leads to the method of ref.~2 for
deriving various RR's for the $SU(2)$ CGc's. The algorithm of
the method can be described as follows. The starting point is
to consider the matrix element
$$
x = < n_1 n_2 n_3 n_4 \vert X \vert j : \mu m \kappa>
\eqno (53)
$$
where $X = {\cal J}_{ \pm }$,
      $   {\Lambda}_{ \pm }$ or
      $    {\cal K}_{ \pm }$. The action of the operator $X$ on
the vector
$\vert j : \mu m \kappa> $ is controlled by
$$
\eqalign{
X\vert j : \mu  m  \kappa > =
( & \sqrt { (j      \mp \mu) (j      \pm \mu +   1) } \quad {\rm or} \cr
  & \sqrt { (j      \mp   m) (j      \pm   m +   1) } \quad {\rm or} \cr
  & \sqrt { (\kappa \mp   j) (\kappa \pm   j \pm 1) } \, ) \, \,
 \vert j : \mu' m' \kappa'> \cr
}
\eqno (54)
$$
where $(\mu'    m'    \kappa') =
       (\mu\pm1 m     \kappa    )$ or
      $(\mu     m\pm1 \kappa    )$ or
      $(\mu     m     \kappa\pm1)$ according to whether as
$X = {\cal J}_{\pm}$ or ${\Lambda}_{\pm}$ or ${\cal K}_{\pm}$.
The resulting scalar product can then be transformed into a
CGc in the $su^{\cal  J}(2)$
        or $su^{\Lambda}(2)$ notation thanks to (51). This provides us with
a first expression of $x$. On the other side, we can calculate
$x$ starting from $< n_1 n_2 n_3 n_4 \vert X$, by using
the boson realization of $X$ and by making use of (53) and (51). This
leads to a new expression of $x$ involving two $SU(2)$ CGc's.
By equating the two expressions obtained for $x$, we get a
three-term RR for the $SU(2)$ CGc's. Along the same vein, other
RR's can be derived by replacing $X$ by a more involved
operator (see ref.~2).

\sb

\noindent {\bf 3.2. The Case $q \ne 1$}

\sb

We now go to a quantum algebra context by replacing the $a$'s
and $b$'s by $q$-bosons. We are thus led to a $q$-oscillator
in four dimensions. For the sake of generality, it would be
interesting to consider the case of a $q$-oscillator in $mn$
dimensions. The Lie groups in the chains (38) would then be replaced
by the corresponding quantum algebras. The existence of
complementary algebras within the so obtained chains of quantum
algebras should be useful. In this direction, it was proved in
ref.~15 that a $q$-analogue $(u_q(m),u_q(n))$ exists, i.e.,
the quantum algebras $u_q(m)$ and $u_q(n)$ are complementary in
the Fock space associated to $mn$ pairs of $q$-boson operators. The
situation is less evident for the couple
$(sp_q(2m,\gr),so_q(n))$. Fortunately, we are interested here
with the case of a $q$-oscillator in $mn=4$ dimensions. In this
case, eq.~(39) can be extended in a quantum algebra context with
three complementary algebras, viz., $su^{\cal J}_q(2)$,
$su^{\Lambda}_q(2)$ and $su^{\cal K}_q(1,1)$.

Indeed, it was
shown in ref.~15 that two complementary algebras of the
$su_q(2)$ type exist~: there are the $q$-analogues of
$su^{\cal J}(2)$ and $su^{\Lambda}(2)$. More precisely, the
operators
$$
\eqalign{
{\cal J}_3 & = {1 \over 2} (N_1 + N_2 - N_3 - N_4) \cr
{\cal J}_+ & = a^+_+ b_+ q^{ {1\over2}(N_2-N_4)}
             + a^+_- b_- q^{-{1\over2}(N_1-N_3)}   \cr
{\cal J}_- & = b^+_+ a_+ q^{ {1\over2}(N_2-N_4)}
             + b^+_- a_- q^{-{1\over2}(N_1-N_3)}   \cr
}
\eqno (55)
$$
and
$$
\eqalign{
{\Lambda}_3 & = {1 \over 2} (N_1 - N_2 + N_3 - N_4)
\cr
{\Lambda}_+ & = a^+_+ a_- q^{ {1\over2}(N_3-N_4)}
              + b^+_+ b_- q^{-{1\over2}(N_1-N_2)}
\cr
{\Lambda}_- & = a^+_- a_+ q^{ {1\over2}(N_3-N_4)}
              + b^+_- b_+ q^{-{1\over2}(N_1-N_2)}
\cr
}
\eqno (56)
$$
generate the quantum algebras $su^{\cal J}_q(2)$ and
$su^{\Lambda}_q(2)$, respectively. Each of these algebras can
be completed to an algebra of type $u_q(2)$ owing to the
operator $N = N_1 + N_2 + N_3 + N_4$. In addition to $N$, a
more interesting invariant of $su^{\cal J}_q(2)$ and
$su^{\Lambda}_q(2)$ can be found~: the operator
$$
[a^+_+ b^+_- q^{ {1\over2}(N_2+N_3)}
             - q^{-1} a^+_- b^+_+ q^{-{1\over2}(N_1+N_4)}     ]q^{1\over2}
$$
is invariant with respect to $su^{\cal J}_q(2)$ and
$su^{\Lambda}_q(2)$. The latter expression plus eq.~(37) suggest
that $su^{\cal K}_q(1,1)$ is spanned by
$$
\eqalign{
{\cal K}_3 & = {1 \over 2} (N_1 + N_2 + N_3 + N_4) + 1                    \cr
{\cal K}_+ & =[a^+_+ b^+_- q^{ {1\over2}(N_2+N_3)}
             - q^{-1} a^+_- b^+_+ q^{-{1\over2}(N_1+N_4)}     ]q^{1\over2}\cr
{\cal K}_- & =[a_+   b_-   q^{ {1\over2}(N_2+N_3)}
             - q^{-1} a_-   b_+   q^{-{1\over2}(N_1+N_4)}     ]q^{1\over2}\cr
}
\eqno (57)
$$
It can be effectively checked that the algebras
$su^{\cal J}_q(2)$,
$su^{\Lambda}_q(2)$ and $su^{\cal K}_q(1,1)$ commute.
Furthermore, the invariants
$$
\eqalign{
C_2(su^{\cal J}_q(2))  &= {\cal J}_- {\cal J}_+ + [{\cal J}_3] [{\cal J}_3 + 1]
\cr
C_2(su^{\Lambda}_q(2)) &= {\Lambda}_- {\Lambda}_+
                        + [{\Lambda}_3] [{\Lambda}_3 + 1]
\cr
C_2(su^{\cal K}_q(1,1))&=-{\cal K}_+ {\cal K}_- + [{\cal K}_3] [{\cal K}_3 - 1]
\cr
}
\eqno (58)
$$
have the same eigenvalues $[j][j+1]$. Hence, the quantum algebras
$su^{\cal J}_q(2)$,
$su^{\Lambda}_q(2)$ and $su^{\cal K}_q(1,1)$ are complementary
and their irrep's can be labelled by the common number $j$.

As a conclusion, the complementarity relations for the
ordinary oscillator in four dimensions can be extended to its
$q$-analogue so that the algorithm, described in section 3.1, for
producing relations between CGc's of the group $SU(2)$ can be
extended to the quantum algebra $su_q(2)$.

Of course, in order to apply the algorithm, we have to be
careful with the similarities and differences between the cases
$q=1$ and $q \ne 1$. In this respect, the following prescriptions
should be taken into account. Equations (33) and (34) have to be
replaced by$^4$
$$
\eqalign{
\vert j_1 m_1> =&{ 1 \over \sqrt{[j_1+m_1]![j_1-m_1]!} } \>
(a^+_+)^{j_1+m_1}
(a^+_-)^{j_1-m_1} \> \vert 00>
\cr
\vert j_2 m_2> =&{ 1 \over \sqrt{[j_2+m_2]![j_2-m_2]!} } \>
(b^+_+)^{j_2+m_2}
(b^+_-)^{j_2-m_2} \> \vert 00>
\cr
}
\eqno (59)
$$
Equations (44)-(46) and (49)-(51) conserve their sense when $q \ne 1$.
Equation  (47) must be changed into
$$
\vert j_1 j_2 jm >_q = \sum_{m_1m_2} (j_1 j_2 m_1 m_2 \vert jm)_q \;
\vert j_1 m_1 > \otimes \> \vert j_2 m_2>
\eqno (60)
$$
Equation (54) should be modified according to eqs.~(8) and (29).

As an example, eq.~(52) can be reinterpreted as a Regge
symmetry property for $su_q(2)$. As another example, we can
find the $q$-analogues, to be reported elsewhere, of the
three-term RR's of ref.~2. Similar results, were obtained by
Nomura$^{11,12}$ and Kachurik and Klimyk$^{16}$
by following different routes. It
should be noted that our algorithm is more powerful since
repeated action of the $q$-deformed operators ${\cal J}_{\pm}$,
${\Lambda}_{\pm}$ and ${\cal K}_{\pm}$ allows us to obtain more
complicated RR's (as, for instance, four-term RR's). In
addition, the $q$-deformation of the
inner product (51) can be interpreted also as a
CGc for $su^{\cal K}_q(1,1)$. Therefore, our algorithm
permits to derive relations connecting CGc's for $su_q(2)$ and
$su_q(1,1)$.

\sb
\sd

\noindent {\bf 4. PERSPECTIVE}

\sb

To close this paper, we would like to point out that
developments similar to the ones in this work are presently
under study, by the authors, for the two-parameter quantum algebra $u_{qp}(2)$.
Such an algebra is spanned by the four
operators $J_-$, $J_3$, $J$ and $J_+$ satisfying the
commutation relations$^{17}$
$$
\eqalign{
[J  , J_3] \;&= \; 0        \qquad
[J  , J_+] \; = \; 0        \qquad
[J  , J_-] \; = \; 0        \cr
[J_3, J_-] \;&= \; - \; J_- \qquad
[J_3, J_+] \; = \; + \; J_+ \qquad
[J_+, J_-] \; = \; (qp)^{J - J_3} \> [2J_3]
}
\eqno (61)
$$
where $[X]$ is now given by
$$
[X] = {{q^X - p^X} \over {p - q}}
\eqno (62)
$$
The algebra $u_{qp}(2)$ admits the invariant
$$
J^2 \; = \; {1 \over 2} \;
(J_+J_- + J_-J_+) + {{[2]} \over {2}} \; (qp)^{J-J_3} \; [J_3]^2
\eqno (63)
$$
and may be endowed with an Hopf
algebraic structure. We forsee from eq.~(63) that $u_{qp}(2)$
presents more flexibility than $su_{q}(2)$ for physical
applications. In particular, the quantum algebra $u_{qp}(2)$
should be of interest in
two-parameter models for rotational spectroscopy of
super-deformed nuclei.$^{18}$

\sb
\sd

\noindent {\bf 5. APPENDIX~: THE ALGEBRA $so_q(3,2)$}

\sb

The nonvanishing commutators of type $[J,J]$, $[K,K]$ and
$[J,K]$ are
given by eqs.~(2), (10) and (13), respectively. The other
    nonvanishing commutators are as follows.

\sd
\noindent {Commutators $[k,k]$~:}
$$
[k^+_+, k^-_-] \; = \; - \; [2K_3 + 2J_3 - 1] \;
                       - \; [2K_3 + 2J_3 + 1] \;
        \rightarrow \; - \; 4(K_3 + J_3)
$$
$$
[k^+_-, k^-_+] \; = \; - \; [2K_3 - 2J_3 - 1] \;
                       - \; [2K_3 - 2J_3 + 1] \;
        \rightarrow \; - \; 4(K_3 - J_3)
$$
\noindent {Commutators $[J,k]$~:}
$$
  [J_3, k^+_+] \;  = \;   k^+_+ \quad
  [J_3, k^+_-] \;  = \; - k^+_- \quad
  [J_3, k^-_-] \;  = \; - k^-_- \quad
  [J_3, k_+^-] \;  = \;   k^-_+
$$
$$\matrix{
    [J_+,k^+_-] & =
 &K_+ ( [ K_3 - J_3 + {3 \over 2} ] &-&
        [ K_3 - J_3 - {1 \over 2} ] )
 &\rightarrow &+ \; 2K_+\cr
    [J_+,k^-_-] & =
 &K_- ( [ K_3 + J_3 + {1 \over 2} ] &-&
        [ K_3 + J_3 - {3 \over 2} ] )
 &\rightarrow &+ \; 2K_-\cr
    [J_-,k^+_+] & =
 &K_+ ( [ K_3 + J_3 - {1 \over 2} ] &-&
        [ K_3 + J_3 + {3 \over 2} ] )
 &\rightarrow &- \; 2K_+\cr
    [J_-,k^-_+] & =
 &K_- ( [ K_3 - J_3 - {3 \over 2} ] &-&
        [ K_3 - J_3 + {1 \over 2} ] )
 &\rightarrow &- \; 2K_-\cr
}$$
\noindent {Commutators $[K,k]$~:}
$$
  [K_3, k^+_+] =   k^+_+ \quad
  [K_3, k^+_-] =   k^+_- \quad
  [K_3, k^-_-] = - k^-_- \quad
  [K_3, k_+^-] = - k^-_+
$$
$$
\matrix{
    [K_+,k^-_-] & =
 &J_- ( [ K_3 + J_3 + {1 \over 2} ] &-&
        [ K_3 + J_3 - {3 \over 2} ] )
 &\rightarrow &+ \; 2J_-\cr
    [K_+,k^-_+] & =
 &J_+ ( [ K_3 - J_3 - {3 \over 2} ] &-&
        [ K_3 - J_3 + {1 \over 2} ] )
 &\rightarrow &- \; 2J_+\cr
    [K_-,k^+_+] & =
 &J_+ ( [ K_3 + J_3 - {1 \over 2} ] &-&
        [ K_3 + J_3 + {3 \over 2} ] )
 &\rightarrow &- \; 2J_+\cr
    [K_-,k^+_-] & =
 &J_- ( [ K_3 - J_3 + {3 \over 2} ] &-&
        [ K_3 - J_3 - {1 \over 2} ] )
 &\rightarrow &+ \; 2J_-\cr
}
$$
The arrows $\rightarrow$ indicate limits when $q$ goes to 1.

\sb
\sd

\noindent {\bf 6. REFERENCES}

\sb

\item{1.} L.C. Biedenharn and J.D. Louck, {\it in}:
Encyclopedia of Mathematics and its Applications, G.-C. Rota,
ed., Addison-Wesley, Reading, Massachusetts (1981) vols. 8 and
9.

\item{2.} M. Kibler and G. Grenet, On the $SU_2$ unit tensor,
J. Math. Phys. 21:422 (1980).

\item{3.} Yu.F. Smirnov, V.N. Tolsto\u \i~and Yu.I. Kharitonov,
Method of projection operators and the $q$ analog of the
quantum theory of angular momentum. Clebsch-Gordan coefficients
and irreducible tensor operators, Sov. J. Nucl. Phys. 53:593 (1991).

\item{4.} L.C. Biedenharn, The quantum group $SU_q(2)$ and a
$q$-analogue of the boson operators, J.~Phys.~A 22:L873 (1989).

\item{5.} A.J. Macfarlane, On $q$-analogues of the quantum
harmonic oscillator and the quantum group $SU(2)_q$, J.~Phys.~A
22:4581 (1989).

\item{6.} M. Kibler and T. N\'egadi, On quantum groups and
their potential use in mathematical chemistry, J.~Math.~Chem.~11:13 (1992).

\item{7.} P.P. Kulish and E.V. Damaskinsky,
On the $q$ oscillator and the quantum algebra $su_q(1,1)$,
J.~Phys.~A 23:L415 (1990).

\item{8.} J. Lukierski, H. Ruegg, A. Nowicki and V.N. Tolstoy,
$q$-deformation of Poincar\'e algebra, Phys. Lett. B 264:331
(1991).

\item{9.} M. Moshinsky and C. Quesne, Noninvariance groups in
the second-quantization picture and their applications,
J. Math. Phys. 11:1631 (1970).

\item{10.} L.C. Biedenharn and M. Tarlini, On $q$-tensor
operators for quantum groups, Lett.~Math.~Phys.~20:271 (1990).

\item{11.} M. Nomura, Recursion relations for the Clebsch-Gordan
coefficient of quantum group $SU_q(2)$, J.~Phys.~Soc.~Jpn.~59:1954 (1990).

\item{12.} M. Nomura, A Jordan-Schwinger representation of
quadratic relations for $SU_q(2)$ operators and of the $q$-analog
Wigner-Eckart theorem, J.~Phys.~Soc.~Jpn.~59:2345 (1990).

\item{13.} C. Quesne, $q$-bosons and irreducible tensors for
$q$-algebras, {\it in}: Symmetries in Science VI, B. Gruber,
ed., Plenum Press, New York (1993).

\item{14.} G.F. Filippov, V.I. Ovcharenko and Yu.F. Smirnov,
Microscopic Theory of
Collective Excitations of Nuclei [in Russian], Naukova Dumka, Kiev (1981).

\item{15.} Yu.F. Smirnov and V.N. Tolstoy, {\it in}: Group
Theory and Special Symmetries in Nuclear Physics, J.P. Draayer
and J. J\"anecke, eds., World Scientific, Singapore (1992) p.~375.

\item{16.} I.I. Kachurik and A.U. Klimyk, On Racah coefficients
of the quantum algebra $U_q(su_2)$, J. Phys. A 23:2717 (1990).

\item{17.} M. Kibler, Introduction to quantum algebras,
IPN de Lyon preprint LYCEN 9234 (1992).

\item{18.} J. Meyer, Yu.F. Smirnov and M. Kibler, work in progress.

\bye